\documentclass[12pt]{article} 

\usepackage[dvips]{graphicx}
\usepackage{xspace}
\usepackage{amssymb,amsfonts,amsthm}
\usepackage{cite}
\newcommand{\bd}{\begin{definition}}                
\newcommand{\ed}{\end{definition}}                  
\newcommand{\bc}{\begin{corollary}}                 
\newcommand{\ec}{\end{corollary}}                   
\newcommand{\bl}{\begin{lemma}}                     
\newcommand{\el}{\end{lemma}}                       
\newcommand{\bp}{\begin{proposition}}            
\newcommand{\ep}{\end{proposition}}                
\newcommand{\bere}{\begin{remark}}                  
\newcommand{\ere}{\end{remark}}                     

\newcommand{\bt}{\begin{theorem}}
\newcommand{\et}{\end{theorem}}

\newcommand{\be}{\begin{equation}}
\newcommand{\ee}{\end{equation}}

\newcommand{\bit}{\begin{itemize}}
\newcommand{\eit}{\end{itemize}}

\newcommand{\R}{\ensuremath{\mathbb{R}}\xspace}     
\newcommand{\abs}[1]{\ensuremath%
            {\vert#1\vert}\xspace}                     

\newcommand{\ptl}{\ensuremath{\omega}\xspace}
\newcommand{\eh}{\ensuremath{F}\xspace}

\newcommand{\dd}{{\rm d}}
\newcommand{\p}{\partial}

\newtheorem{theorem}{Theorem}[section]
\newtheorem{corollary}[theorem]{Corollary}
\newtheorem{lemma}[theorem]{Lemma}
\newtheorem{proposition}[theorem]{Proposition}
\newtheorem{definition}[theorem]{Definition}
\newtheorem{remark}[theorem]{Remark}

\begin{document}

\title{Maximizing curves for the charged-particle action in globally hyperbolic spacetimes}

\author{E. Minguzzi \\  \small{ {\em Departamento de Matem\'aticas, Universidad de Salamanca,
}} \\ \small{ {\em Plaza de la Merced 1,
 E-37008 Salamanca, Spain}}\\ \small{ {\em  and INFN, Piazza dei Caprettari 70, I-00186
Roma, Italy}}\\
\small{minguzzi@usal.es}}

\date{}



\maketitle
\begin{abstract}
In a globally hyperbolic spacetime any pair of chronologically
related events admits a connecting geodesic. We present two
theorems which prove that, more generally, under weak assumptions,
given a charge-to-mass ratio there is always a connecting solution
of the Lorentz force equation having that ratio. A geometrical
interpretation of the charged-particle action is  given which
shows that the constructed solutions are maximizing.
\end{abstract}


\section{Introduction}
Over the last years there has been considerable interest in the
connectability of spacetime events through solutions of the
Lorentz force equation
\cite{bartolo99,bartolo00,bartolo01,bartolo02,bartolo03,bartolo04,antonacci00,
caponio02,caponio02b,caponio04,caponio04c,mirenghi02,mirenghi04,piccione04}.
The works on this topic are  mainly concerned with the problem of
determining  whether a charged particle can reach an event from
another on its past. In mathematical terms the problems is
\\

\noindent Given a globally hyperbolic spacetime $M$, an event
$x_{0}$ and a second event $x_{1} \in I^{+}(x_{0})$ determine
whether  the Lorentz force equation (cf.
\cite{jackson75,misner73})
\begin{equation} \label{lorentz}
 D_s\!\left(\frac{\dd x}{\dd s}\right)=\frac{q}{m}\hat F(x)\left[\frac{\dd x}{\dd
 s}\right],
\end{equation}
admits connecting solutions. Here $x=x(s)$ is the worldline of the
particle parameterized with respect to the proper length,
$\frac{\dd x}{\dd s}$ is the $4$-velocity, $D_s\!\left(\frac{\dd
x}{\dd s}\right)$ is the covariant derivative of $\frac{\dd x}{\dd
s}$ along $x(s)$ associated to the Levi-Civita connection of $g$,
and $\hat F(x)[\cdot]$ is the linear map  on $T_x M$ obtained
raising the first index of $F$. Conventions are such that $c=1$
and the metric $g$ has signature $(+ - - -)$.
\\

Actually, the references cited above studied the connectability of
spacetime through solutions of the equation
\begin{equation} \label{lorentz2}
 D_{\lambda} \left(\frac{\dd x}{\dd
\lambda}\right)=Q\hat F(x)\left[\frac{\dd x}{\dd \lambda}\right],
\end{equation}
where $\lambda$ is a generic parameter. Although apparently of the
same form of the Lorentz force equation this equation is
considerably weaker. Every solution of the Lorentz force equation
is, suitably parametrized and independently of the charge-to-mass
ratio, a solution of this last equation. As a consequence, as the
space of solutions of this last equation is infinitely larger than
the one of the Lorentz force equation, its is easier to find
connecting solutions for it. What makes the Lorentz force equation
more restrictive is the condition that the 4-velocity should be
{\em a priori} normalized while the same condition on $\frac{\dd
x}{\dd \lambda}$ is not imposed.

Substantial progress was made using a Kaluza-Klein approach
\cite{caponio01,minguzzi03b,caponio03}. The only connectability
results up to now available on the  Lorentz force equation
(\ref{lorentz}) are indeed those in \cite{minguzzi03b,caponio03}.
We shall here review the basic ideas behind these two references.
We shall skip the more technical proofs. Instead we shall focus on
the geometrical meaning of some results that has not been
previously pointed out.

The problem we are going to study is therefore that of
generalizing the Avez-Seifert theorem \cite{avez63,seifert67}
\begin{theorem}
In a globally hyperbolic spacetime let $x_{1} \in J^{+}(x_{0})$
then there is a geodesic connecting the two events.
\end{theorem} \noindent
to each choice of $q/m \ne 0$ (the Avez-Seifert theorem
corresponds to the case $q/m=0$). We restrict to the case $x_{1}
\in I^{+}(x_{0})$ since only timelike connecting solutions can be
interpreted as charged particles.

\begin{figure}[!ht]
\centering
\includegraphics[width=8cm]{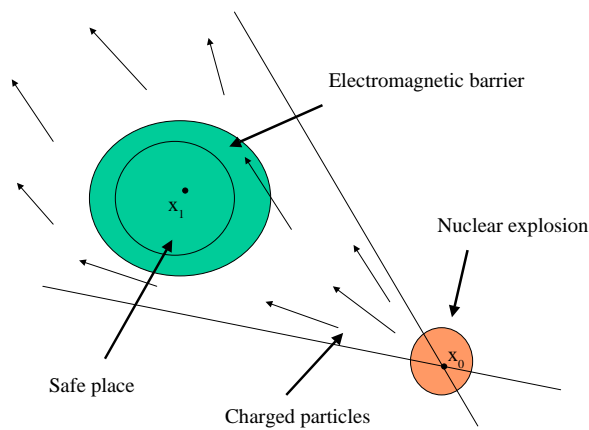}
\caption{Is it possible to screen charged particles with an
electromagnetic barrier?} \label{figura1}
\end{figure}

Before going on we would like to point out that an affirmative
solution to the problem would imply that it is impossible to
construct an ``electromagnetic barrier''. Eventually we shall see
that the problem admits indeed an affirmative answer. Since it is
impossible to isolate an event from charged particles emitted from
an event in its past it is also impossible to isolate an entire
spacetime region. Figure \ref{figura1} dramatically shows the
problem of protecting ourselves from the charged particles emitted
from a nuclear explosion by means of a suitable electromagnetic
field around us. No electromagnetic field can protect us from the
emitted particles.
A consequence of our results is that  one has to place walls
between the nuclear explosion and the observer, i.e. one has to
involve the absorbtion process of quantum  physics to screen the
particles.

The problem considered can be formulated as a variational one. Let
$\gamma: [0,1] \to M$, be a  $C^{1}$ curve, if timelike solutions
of the Lorentz force equation connecting $x_{0}$ and $x_{1}$
exist, then they are extremals of
 \[
I_{x_0,x_1}[\gamma]=\int _{\gamma} (\dd s+\frac{q}{m}  \omega),
\qquad \gamma(0)=x_{0}, \ \gamma(1)=x_{1}.
\]
where $\dd s= \sqrt{g_{\mu \nu} \dot x^{\mu} \dot x^{\nu}} \dd
\lambda$ and $\omega$ is the 1-form potential over $M$. The
electromagnetic field is
\begin{equation}
F=\dd \omega,
\end{equation}
thus we are here considering only the case of exact
electromagnetic field. This is not a strong requirement. Up to now
there has been no experimental evidence of non-exact
electromagnetic fields on spacetime although a lot of effort was
devoted to the search of magnetic monopoles. Although
mathematicians are used to consider the electromagnetic field as a
closed 2-form, as observation is concerned, the electromagnetic
field is an exact 2-form. In particular this implies that the
electromagnetic field can be regarded as the curvature of a {\em
necessarily  trivial} $(\mathbb{R}, +)$ bundle over $M$.

Our problem of connectability can be stated as follows: Does the
action functional above have  timelike extremals?, and maximizing
extremals in a suitable set?

\section{Basic definitions}
\begin{figure}[ht] \centering
\includegraphics[width=4cm]{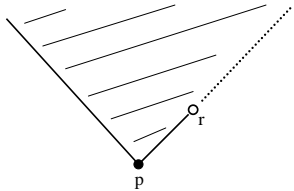}
\caption{$J^{+}(p)$ is not necessarily closed.} \label{figura3}
\end{figure}

We need some basic facts and definitions that we recall in this
section. The spacetime $(M,g)$ is said to be time-orientable if it
has a timelike continuous vector field $v$. We assume $M$
time-orientable and make a choice of past and future. We recall
that a $C^{1}$ curve $\gamma: \mathbb{R} \to M$ is said
\begin{itemize}
\item timelike if: $g(\dot x,\dot x)>0$,
\item non-spacelike if: $g(\dot x,\dot x) \ge 0$,
\item null if : $g(\dot x,\dot x) = 0$,
\item future-directed if: $g(v,\dot x)>0$.
\end{itemize}
for any $\lambda \in \mathbb{R}$. We shall always consider
future-directed curves. We can now recall the definitions of the
future sets
\begin{itemize}
\item Chronological future: $I^{+}(p)=\{q \in$ $M$ that are
connected to $p$ through a future-directed timelike $C^{1}$ curve
$\}$
\item Causal future: $J^{+}(p)=\{q \in$ $M$ that are
connected to $p$ through a future-directed non-spacelike $C^{1}$
curve $\}$
\end{itemize}
The following properties hold. A proof can be found, for instance,
in \cite{hawking73,oneill83}.
\begin{itemize}
\item $I^{+}(p) \subset J^{+}(p)$
\item $I^{+}(p)$ is open
\item $\bar I^{+}(p) = \bar J^{+}(p)$
\item $\dot I^{+}(p) = \dot J^{+}(p)$
\end{itemize}
where with $\dot K= \bar K \cap \bar{(M-K)}$ we denoted  the
boundary of $K$. Note that $J^{+}(p)$ is not necessarily closed.
Figure \ref{figura3} shows an example where an event has been
removed from Minkowski spacetime.
We shall see below that this spacetime is not  globally
hyperbolic.

The future horismos of $p \in M$ is the set
\[
E^{+}(p)=J^{+}(p)-I^{+}(p)
\]
The following theorem holds \cite{hawking73}

\begin{figure}[!ht]
\centering
\includegraphics[width=4cm]{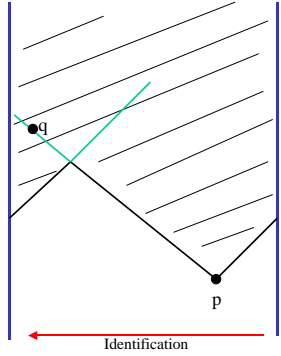}
\caption{There can be events in null geodesics from $p$ that do
not belong to $E^{+}(p)$.} \label{figura2}
\end{figure}

\begin{theorem} Any non-spacelike curve between two events which is
not a null geodesic can be deformed into a timelike curve between
the same points.
\end{theorem}
Then
\begin{corollary}
{\em Any point  $ q \in E^{+}(p)$ is connected to $p$ by a null
geodesic}\vspace{10pt}
\end{corollary}
But the converse is not true. There can be events in null
geodesics from $p$ that do not belong to $E^{+}(p)$, see figure
\ref{figura2} where an example is given in a cylindrical
spacetime.

We now recall that a curve $\gamma$ has {\em future endpoint} $p$
if for any neighborhood $U$ of $p$ there is a $\lambda_{1}$ such
that $\gamma(\lambda) \in U$ for $\lambda > \lambda_{1}$. A curve
is said {\em future-inextendible} if $\gamma$ has no future
endpoints. A curve is said {\em inextendible}  if $\gamma$ is both
past and future-inextendible. A  {\em Cauchy surface} is a
spacelike hypersurface which every non-spacelike inextendible
curve intersects exactly once. We are ready to recall a basic
definition that we shall use throughout the work. A spacetime $M$
is said {\em globally hyperbolic}  if it admits a Cauchy
hypersurface. We also recall that a spacetime $M$ satisfies the
{\em strong causality condition} if any event $p$ has a
neighborhood that no non-spacelike curve intersects more than
once. The following remarkable results holds \cite{hawking73}

\begin{theorem}
 The Lorentzian manifold $M$ is globally hyperbolic iff the strong causality
condition holds and for any pair of events $J^{+}(p) \cap
J^{-}(q)$ is compact.
\end{theorem}
From this it follows \cite[Proposition 6.6.1]{hawking73} a result
that will be central in our study
\begin{corollary}
If $M$ is globally hyperbolic then
\[
J^{+}(p)=\bar J^{+}(p), \qquad  E^{+}(p)=\dot J^{+}(p)=\dot
I^{+}(p)
\]
\end{corollary}
Thus by a previous observation
\begin{corollary} \label{cor}
If $M$ is globally hyperbolic then events in $\dot I^{+}(p)$ are
connected to $p$ by null geodesics.
\end{corollary}

\section{Kaluza-Klein spacetime}
Consider a trivial 5-dimensional principal bundle $P=M \times
\mathbb{R}$ of structure group $T_1=(\mathbb{R},+)$, and
projection $\pi : P \to M$ such that $(m,y) \to m$. Here $y$ is
the (dimensionless) coordinate over $\mathbb{R}$. We can define
the real-valued
 connection 1-form $\tilde\omega$  on $P$:
\[
\tilde\omega=(\dd y + \beta \ptl).
\]
The coefficient $\beta$ is introduced for dimensional reasons but
will be otherwise arbitrary. Over $P$ we introduce  the
Kaluza-Klein metric
\begin{equation} \label{kk}
\tilde{g}=g-a^{2}(\dd y + \beta \omega)^{2}
\end{equation}
The constant $a$ represents the scale factor of the extra
dimension and will be later fixed.

\begin{remark}
The scale factor $a$ is interpreted in the $U(1)$ version as the
radius of the extra dimension. The field equations for general
relativity in 5-dimensions are seen in 4-dimensional spacetime, as
4-dimensional general relativity plus electromagnetism. In fact
the 5d Einstein-Hilbert Lagrangian is, if one chooses
$\beta=\sqrt{16\pi G}/a$ with $G$ the Newton constant,
\[
\tilde{R}=R+\frac{16\pi G}{4}F_{\mu \nu} F^{\mu \nu}.
\]
as it should be in order to reproduce the correct coupling between
electromagnetism and gravity. Here we use the Kaluza-Klein
spacetime only as a technical tool. We never use this physical
interpretation so we do not need to relate $\beta$ and $a$ as
above.
\end{remark}

\begin{figure}[!ht]
\centering
\includegraphics[width=6cm]{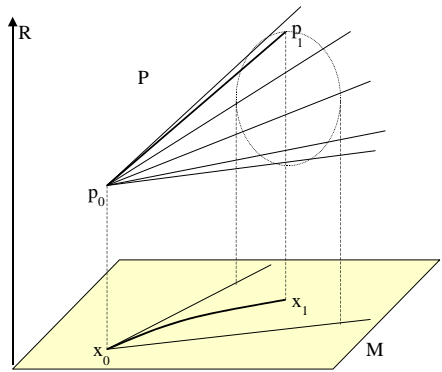}
\caption{Constructing the solution using Kaluza-Klein.}
\label{figura5}
\end{figure}

The projection of a 5d geodesic is a 4d solution to the Lorentz
force equation for a suitable charge-to-mass ratio. Indeed
geodesics $z(\lambda)=(x(\lambda),y(\lambda))$ are extremals of
the functional
\[
S=\int_{0}^{1} \frac{1}{2}\tilde{g}[\dot z(\lambda),\dot
z(\lambda)]\dd \lambda,
\]
where with a dot we have denoted derivation with respect to
$\lambda$. The Lagrangian is independent of $y$ thus the vertical
conjugated momentum
\[
p=\frac{\p L}{\p \dot{y}}=-a^{2}(\dot{y} + \beta\ptl[\dot{x}]),
\]
is conserved. The other Euler-Lagrange equation is
\begin{equation} \label{pro}
\label{x} D_{\lambda} \dot x=p \beta \hat{{F}}[\dot x],
\end{equation}
which is quite similar to the Lorentz force equation. The previous
equation implies that $g(\dot x,\dot x)=C^{2}$ is conserved
(contract equation (\ref{pro}) with $\dot x$), and since $z$ is a
geodesic $\tilde{g}(\dot z, \dot z)$ is conserved too. We have
that the conserved quantities are related by
\[
\tilde{g}(\dot z, \dot z)=C^{2}-\frac{p^{2}}{a^{2}} .
\]
There follows
\begin{remark} \label{boh}
If $z$ is a null geodesic then $C^{2}  = p^{2}/a^{2}$ and in this
case $x=\pi(z)$ is timelike iff $p\neq 0$.
\end{remark}
Suppose we want to study the  connectability of spacetime through
solutions of the Lorentz force equation of charge-to-mass ratio
$q/m$. Our strategy in order to find such connecting solutions is
based on the following
\begin{remark}
Consider on $P$ the Kaluza-Klein metric with $a= \frac{1}{\beta}
\vert \frac{q}{m} \vert$. Let $z$ be a null geodesic on $P$. If
$x=\pi(z)$ is timelike then it satisfies the Lorentz force
equation of charge-to-mass ratio $\pm \vert q/m\vert$ where the
sign is that of $p$.
\end{remark}
This remark can be proved easily by substituting $\dd \lambda= \dd
s /C$ in (\ref{pro}). Our strategy will be therefore the
following. Chosen $a$ as above and a point $p_{0} \in
\pi^{-1}(x_{0})$ we look for two null geodesics starting from
$p_{0}$ and ending ($\lambda=1$) on  $x_{1}$'s fiber (see figure
\ref{figura5}).
 If their projections are timelike and the $p$ sign is
respectively $+$ and $-$ then their projections solve
affirmatively the problem of connectability for the charge-to-mass
ratios $+\vert \frac{q}{m} \vert$ and $-\vert \frac{q}{m} \vert$.

\begin{figure}[!hb]
\centering
\includegraphics[width=6cm]{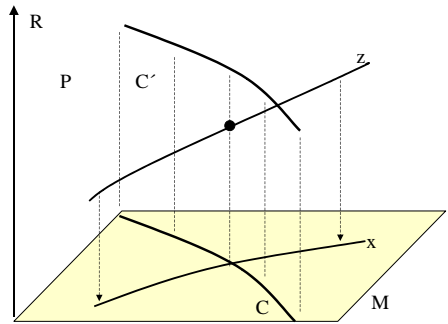}
\caption{$P$ is globally hyperbolic.} \label{figura6}
\end{figure}

Let us now show that the two null geodesics above indeed exist.
The set $J^{+}(p_{0}) \cap \pi^{-1}(x_{1})$  is a compact subset
of $x_{1}$'s fiber (real line). The maximum $\bar p$ and the
minimum $\hat p$ of this set belong to $\dot J^{+}(p_0)=\dot
I^{+}(p_0)$. Then by corollary \ref{cor}, if we shown that the
Kaluza-Klein spacetime is itself globally hyperbolic, there are
null geodesics on $P$ that reach them starting from $p_{0}$. Their
projections if timelike can be shown to have opposite signs
\cite{minguzzi03b} of $p$ as required. We are therefore left only
with two open questions
\begin{itemize}
\item If $M$ is globally hyperbolic is $P$ globally hyperbolic?
\item What can we say if the projection of the null geodesics so constructed in $P$ are null curves?
\end{itemize}
It is not difficult to show that the first question admits a
positive answer \cite{caponio01,caponio03}. The proof has a clear
geometrical meaning shown in figure \ref{figura6}.
If $M$ is globally hyperbolic then it has a Cauchy hypersurface
$C$. The proof goes on showing that  $\pi^{-1}(C)$ is a Cauchy
hypersurface for $P$. Indeed if  $z$  is an inextendible timelike
curve in $P$ its projection $x$ is timelike and inextendible in
$M$ (this is the only technical point in the proof) and thus
intersects $C$ in some point. Therefore $z$  intersects
$\pi^{-1}(C)$ at some point i.e. $P$ is globally hyperbolic.

The answer to the second question is that by Eq. (\ref{pro}) and
remark \ref{boh}, $p=0$ and therefore the projection is a null
geodesic. If there are no connecting null geodesics between the
events $x_{0}$ and $x_{1}$ this case is excluded. A typical
example is Minkowski spacetime: if $x_{1} \in I^{+}(x_0)$ then
there is no null geodesics connecting the events.

We are ready to state the first result obtained but first let us
see what can be said about the associated variational problem.

\section{A geometrical interpretation for the action}
In \cite{minguzzi03b} the author found a simple geometrical
interpretation for the functional $I$ which is somewhat
reminiscent of the time of arrival functional in the Fermat
principle of general relativity \cite{kovner90,perlick90}.

Given a (future-directed) causal connecting curve
$\sigma(\lambda): [0,1] \to M$,  consider two lifts of the curve
(dependent on the sign) in   null curves $\tilde\sigma^{\pm}(\lambda)$
of $P$ starting at $p_{0}=(x_{0}, y_{0})$, through the following
definition
\begin{equation}
\tilde{\sigma}^{\pm}(\lambda)=(\sigma(\lambda), y^{\pm}(\lambda))=
\left(\sigma(\lambda),\, y_0 \mp \frac{1}{a} \int_{ \sigma(0)}^{\
\sigma(\lambda)} \dd s-\beta \int_{ \sigma(0) }^{\
\sigma(\lambda)}\omega \right).
\end{equation}
These are the curve ``light lifts'' to be distinguished from the
usual ``horizontal lift".
\begin{figure}[!hb]
\centering
\includegraphics[width=6cm]{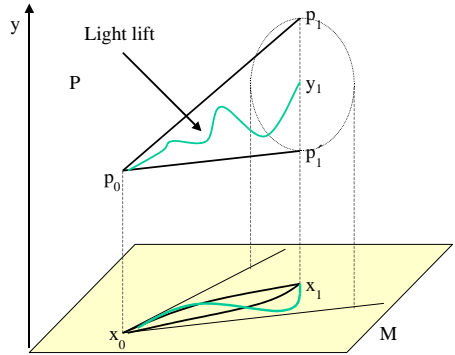}
\caption{The light lift.} \label{figura7}
\end{figure}
The fiber coordinate $y_{1}^{\pm}$ of the final point of this curve
is, essentially, the action functional:
\begin{equation} \label{ciao}
y_{1}^{\pm}=y_{0}\mp \frac{1}{a}\left(\int_{\sigma} \dd s +(\pm \vert
q/m\vert)  \int_{\sigma} \omega \right).
\end{equation}
thus a maximization on the space of $C^{1}$ connecting causal
curves, $\mathcal{N}_{x_0,x_1}$, of the functional
$I_{x_{0},x_{1}}$ relative to the ratio $+ \vert q/m \vert$ (resp.
$- \vert q/m \vert$), corresponds to  a minimization (resp.
maximization) of $y_{1}^{+}(\sigma)$ (resp. $y_{1}^{-}(\sigma)$).
The maximizing curve is exactly the one constructed in the
previous section in order to find a solution to the Lorentz force
equation. In summary we have the following \cite{minguzzi03b}:

\begin{theorem}\label{main}
Let $(M,g)$ be a globally hyperbolic spacetime, $ \ptl $ be a
($C^{2}$) 1-form  on $M$,  and $\eh=\dd \ptl$. Let $x_{1}$ be an
event in the chronological future of $x_{0}$ and $q/m \in \R-\{ 0
\}$ any charge-to-mass ratio.   Let $P=M \times \mathbb{R}$ be the
Kaluza-Klein spacetime of metric (\ref{kk}), with
$a=\frac{1}{\beta} \vert \frac{q}{m} \vert$. Let $p_{0}$ be a
point in the fibre of $x_{0}$ and let the compact set
$J^{+}(p_{0}) \cap \pi^{-1}(x_{1})$ have endpoints
$\bar{p}_{1}=(x_{1},\bar{y}_{1})$,
$\hat{p}_{1}=(x_{1},\hat{y}_{1})$ with $\bar{y}_{1} \ge
\hat{y}_{1}$. Let the curve $\gamma_{0}$ be the projection of a
null geodesic $\bar\gamma$ that connects $p_{0}$ to $\bar{p}_{1}$
if $q/m <0$ or  the projection of a null geodesic $ \hat{\gamma} $
 that connects $p_{0}$ to $\hat{p}_{1}$ if
$q/m>0$.

The null geodesics $\bar\gamma$ and $\hat\gamma$ exist, and the
future-directed causal  curve $\gamma_0$
 connecting $x_0$ and $x_1$
maximizes the functional $I[\gamma](x_0,x_1)$ on the space
$\mathcal{N}_{x_0,x_1}$. Moreover, $\gamma_0$ being the projection
of a null geodesic is everywhere timelike or null. In the former
case, the reparametrization of $\gamma_0$ with respect to proper
time is a solution of the Lorentz force equation (\ref{lorentz});
in the latter case, $\gamma_0$ is a null geodesic.
\end{theorem}

\begin{corollary}
 Let $(M,\eta)$ be the Minkowski spacetime.  Let
$\eh$ be an electromagnetic tensor field (closed 2-form).  Let
$x_{1}$ be an event in the chronological future of $x_{0}$ and
$q/m$ a charge-to-mass ratio, then there exists at least one
future-directed timelike solution to the Lorentz force equation
connecting $x_0$ and $x_1$.
\end{corollary}

\begin{proof}
Since $M$ is contractible $\eh$ is exact. Moreover, in Minkowski
spacetime, if $x_{1} \in I^{+}(x_{0})$ there is no null geodesic
connecting $x_{0}$ with $x_{1}$.
\end{proof}

\section{A complete answer in the classical case}
Unfortunately the previous theorem leaves open the possibility
that $\gamma_{0}$ is a null geodesic. We ask whether it is
possible to discard this possibility. If the charge-to-mass ratio
is smaller in absolute value of a certain real number the answer
is affirmative \cite{caponio03}. Define
\begin{equation}
R=\sup_{x \in \mathcal{T}_{x_0,x_1}} \Big(\frac{ \int_{x} \dd s}{
\sup_{w \in \mathcal{N}_{x_0,x_1}}|\int_{w} \omega - \int_{x}
\omega|}\Big). \label{L}
\end{equation}
where $\mathcal{T}_{x_0,x_1}$ is the space of $C^{1}$ timelike
connecting curves. Note that $R$ does not depend on the gauge
chosen. It can be shown  that $R$ is strictly positive and
moreover the following theorem holds \cite{caponio03}

\begin{theorem} \label{pri}
Let $(M,g)$ be a time-oriented Lorentzian manifold. Let \ptl be a
1-form ($C^{2}$) on $M$ (an electromagnetic potential) and
$\eh=\dd \ptl$ (the electromagnetic tensor field). Assume that $(M
,g)$ is a globally  hyperbolic manifold. Let $x_{1}$ be an event
in the chronological future of $x_{0}$ and let $R$  be defined as
in (\ref{L}), then there exists at least one future-directed
timelike solution to the Lorentz force equation connecting $x_0$
and $x_1$, for any charge-to-mass ratio satisfying $
\abs{\frac{q}{m}}< R$.
\end{theorem}

\begin{proof} Let $z=(x,y)$ be the null geodesic that connects $p_{0}$
with $\bar{p}$ (notations of the previous theorem). From
$p=-a^{2}(\dot{y} + \beta\ptl[\dot{x}])$, integrating and
introducing an arbitrary timelike connecting curve $\sigma$
\begin{eqnarray*}
p&=&-a^{2}(y_{1}-y_{0}+ \beta\int_{x}
\ptl)\\
&=&-a^{2}(y_{1}-y_{0}+\beta\int_{\sigma}
\omega)-a^{2}\beta(\int_{x} \omega - \int_{\sigma} \omega)\\
&=&-a^{2}(y_{1}-y_{1}^{-}[\sigma] + \frac{1}{a}\int_{\sigma} \dd s
)-a^{2}\beta(\int_{x} \omega - \int_{\sigma} \omega)
\end{eqnarray*}
Since $z$ connects $p_0$ with $\bar{p}$, by construction it is
$y_{1}-y_{1}^{-}[\sigma] \ge 0$, thus
\begin{eqnarray*}
p &\le & -a\int_{\sigma} \dd s -a^{2}\beta(\int_{x} \omega -
\int_{\sigma} \omega) \\ &=& - \left(1+a \beta\frac{\int_{x}
\omega - \int_{\sigma} \omega}{\int_{\sigma} \dd s}\right)a
\int_{\sigma} \dd s
\end{eqnarray*}
If \[ a \beta= \vert \frac{q}{m}\vert < \vert\frac{\int_{\sigma}
\dd s}{\int_{x} \omega - \int_{\sigma} \omega}\vert \] for some
$\sigma$ then $p <0$ and the projection of $z$ is timelike as we
want to prove. If $\vert q/m \vert < R$ there is indeed a timelike
curve $\sigma$ such that
\[
\vert \frac{q}{m} \vert <\frac{ \int_{\sigma} \dd s}{ \sup_{w \in
\mathcal{N}_{x_0,x_1}}|\int_{w} \omega - \int_{\sigma} \omega|}
\le \vert \frac{\int_{\sigma} \dd s}{\int_{x} \omega -
\int_{\sigma} \omega} \vert ,
\]
 and therefore $p<0$. An analogous reasoning holds with $z$ connecting $p_{0}$ and $\hat
p$ and leads to a conserved vertical momentum of opposite sign $p
> 0$. This concludes the proof.
\end{proof}

\subsection{Physical meaning of theorem \ref{pri}}
In quantum mechanics a particle behaves as it could experience all
the possible paths (quantum histories) between two events. The
condition $\vert q/m \vert<R$ can be written
\[
\sup_{x \in \mathcal{T}_{x_0,x_1}} \frac{m}{q\Delta \bar
\Phi_{x}}>1
\]
the denominator
\[
q\Delta \bar \Phi_{x}=\frac{ \mathcal{N}_{x_0,x_1}|\int_{w}
q\omega - \int_{x} q\omega| }{\int_{x} \dd s}
\]
 is the maximum mean (with respect to proper time)
variation of energy potential that a particle moving on $x$
experiences with respect to the other alternatives (the curves
$w$). Now, if the particle moves on $x$ it must be
\[
q\Delta \bar \Phi_{x}<m
\]
otherwise there would be pair creations. If the particle moves
without pair creation effects  then there must be at least a curve
such that the previous equation holds
\[
\inf_{x \in \mathcal{T}_{x_0,x_1}}{q\Delta \bar \Phi_{x}}<m
\]
which coincides with the condition of the theorem. We conclude
that the condition $\vert\frac{q}{m}\vert<R$ means {\em classical}
regime. If this condition is not satisfied there can be pair
creation effects (like in the so called Klein paradox) and the
Lorentz force equation is no longer valid in the quantum regime.
If the condition is satisfied we are in a classical regime and the
theorem states that there are indeed connecting solutions to the
Lorentz force equation. Note that in the limit $\omega \to 0$, we
have $R \to \infty$ which means that in a weak electromagnetic
field we are always in a classical regime an the theorem can be
applied (notice that the word {\em weak} here does not mean that
the theorem holds in some approximation or in some limit).

\section{Conclusions}

We have presented two theorems that answer affirmatively to the
existence of connecting solutions to the Lorentz force equation
(\ref{lorentz}). One of them implies that there is a connecting
solution if there is no null connecting geodesic as it happens,
for instance, in Minkowski spacetime. The other shows that even if
this is not the case, if the absolute value of the charge-to-mass
ratio is less than a certain geometrical gauge invariant number $R
> 0$, there are connecting solutions. This last condition has been
shown to correspond to the classical limit, i.e. to the case where
there are no quantum mechanical pair creations effects and it is
therefore completely satisfactory from the physical point of view.
Indeed, in a non-classical (quantum mechanical) regime the very
Lorentz force equation becomes  meaningless. Finally, thanks to a
geometrical interpretation of the charged-particle action the
solutions above have been shown to maximize the action.

\section*{Acknowledgements}
The author thanks E. Caponio and M. S\'anchez Caja for useful
conversations. This work has been  supported by INFN, grant
$\textrm{n}^{\circ}$ 9503/02.


\end{document}